\documentstyle[psfig,nicV]{article}
\begin{document}
%
%
\heading{%
Production of the LiBeB isotopes with\\
time-dependent GCR fluxes\\
%
}
\par\medskip\noindent
%
\author{%
Carlos Abia$^{1}$, Ute Lisenfeld$^{2}$, Jordi Isern$^{3}$
}
\address{
Dpt. F\'\i sica Te\'orica y del Cosmos, Universidad de Granada, Granada (Spain)}
\address{
IRAM, Granada (Spain)}
\address{
Institut d'Estudis Espacials de Catalunya-CSIC, Gran Capit\'a 2-4, Barcelona (Spain)
}
\begin{abstract}

It is well known that no time-independent galactic cosmic ray (GCR) flux model
can account for the observed linear relationship of Be and B abundances 
with metallicity. In this contribution we solve the time-dependent CR propagation 
equation using simple analytic time dependences for the CR 
spectrum at the source, the galactic CR escape-length and gas density evolution. 
The CR fluxes so obtained are used to study the early evolution of the 
LiBeB abundances in the Galaxy.
\end{abstract}

\section{A time-dependent model for the GCR propagation}
Recent observations of unevolved halo stars have confirmed the surprising linear 
relationship of the Be and B abundances with [Fe/H] (\cite{Du}, \cite{Ga}). This 
primary behaviour of Be and B  cannot be understood on the basis of the 
standard model (SM) of LiBeB production in the interstellar medium (ISM) from GCR 
spallation. This is because in the SM the LiBeB production rates are proportional 
to the global metallicity of the ISM, so that a secondary behaviour with [Fe/H] is 
expected. However, the SM assumes a {\it GCR flux constant in time} and equal to that 
currently oberved at the Earth. There are many reasons to believe that the  GCR flux was 
different in early epochs. In this contribution we solve the time-dependent propagation 
equation of CRs in the galaxy in a simple and parametrized way to test whether the SM 
can be compatible with observations when the hypothesis of the CR flux constancy in 
time is relaxed.

We consider a general time-dependent equation for the GCR propagation:
\begin{equation}
{\partial F(E,t)\over\partial t} + {F(E,t)\over\Lambda(E,t)}
-{\partial(w(E,t)F(E,t))\over\partial E} = q(E,t) 
\end{equation}
where $F(E,t)$ is the number density of GCR particles per energy interval as a function 
of energy per nucleon and time, $\Lambda(E,t)$ is the loss time-scale including escape, 
$\Lambda_e(E,t)$, and inelastic nuclear interaction, $\Lambda_n(E,t)$. 
$w(E,t)$ represents the ionization energy losses in the ISM and finally, $q(E,t)$ is the 
source spectrum of GCRs. We make the following assumptions:

\noindent
1) The energy and time dependences of $\Lambda_{e,n}$, $w$ and $q$ are separable:
\begin{equation} 
\Lambda_{e,n}(E,t)=\Lambda_1^{e,n}(E)\Lambda_2^{e,n}(t);
w(E,t)=w_1(E)w_2(t);
q(E,t)=q_1(E)q_2(t)
\end{equation}
2) We assume that GCRs are produced by supernovae (SN) so that $q_2(t)=SNR(t)$, 
the SN rate, taken from a chemical evolution model which reproduces the current SN rates 
in the galaxy and the main characteristics of the solar neighbourhood. The source energy 
spectrum is: 
\begin{equation}
q_1(E)=q_0{(E+E_0)\over[E(E+2E_0)]^{1.5}} 
\end{equation}
with $E_0=938$ MeV. The constant $q_0$ is chosen in such a way that we can 
reproduce the observed current GCR flux at the Earth. 

3) $\Lambda(E,t)$ can be split into two parts
\begin{equation}
{1\over\Lambda(E,t)}={1\over\Lambda_e(E,t)} + {1\over\Lambda_n(E,t)} 
\end{equation}
where $\Lambda_e$ represents the escape time-scale from the galaxy, and $\Lambda_n$ is
the loss time-scale against inelastic nuclear interaction. The energy dependence of 
$\Lambda_e$ is taken from \cite{Gar} and that of $\Lambda_n$ from \cite{Me} assuming a 
gas density of 1 cm$^{-3}$.

4) The energy losses due to ionization in the ISM have been considered according
to \cite{Lo}.

5) The time-dependence of the ionization losses and of the inelastic nuclear interactions 
can be plausibly caused by temporal changes in the gas density. The time-dependence of 
the escape time-scale is more difficult to understand. Here we treat it as a free 
parameter. For both we assume, for the sake of simplicity, exponential functions:
\begin{equation}
\Lambda_2^e(t)=\exp(-t/\tau_{esc}); \Lambda_2^n(t)=\exp(-t/\tau_{gas}); 
w_2(t)=\exp(-t/\tau_{gas})
\end{equation}
where $\tau_{gas}$ is the characteristic e-folding time of gas exhaustion in 
the galaxy and $\tau_{esc}$ is the characteristic time-scale for variations of 
$\Lambda_e$. We assume that $\Lambda(E,t)$ was higher in the past, during the halo 
collapse (t$\leq 2$ Gyr).
\section{The Solution}

Eq. (1) can be solved using the method of characteristics. The solution
can be simplified considerably if $\tau_{gas},\tau_{esc} \gg 10^7$ yr. Plausible 
values for $\tau_{gas}$ and $\tau_{esc}$ are longer than $10^7$ yr, so in that case 
the solution is
\begin{eqnarray}
F(E,t)&=&{1\over w_1(E)w_2(t)} \int^\infty_E dE_0 \, q_1(E_0)\, 
q_2\left(t-\tau_{loss}(E_0,t)\right)\nonumber\\
&&\exp\left(- \int_E^{E_0} {dE' \over w_1(E') \Lambda_1^e(E')
\Lambda_2^e(t)\Lambda_1^n(E')}\right)
\end{eqnarray}
where $\tau_{loss}(E',t)$ is the ionization energy loss time-scale calculated as:
\begin{equation}
\tau_{loss}(E',t)=\int_E^{E'} dE'' {1\over w_1(E'') w_2(t)}
\end{equation}
Then, the rate of accumulation of LiBeB in the ISM is calculated in the usual 
way (see \cite{Fi}). We have considered the following cases:

{\bf Case 1:} The energy dependence of the GCR source spectrum is given 
by Eq. (3) with $q_2(t)$ equal to the SN rate evolution. The remaining  
parameters are constant in time. 

{\bf Case 2:} Case 1 plus $\Lambda_2(t)$ from Eq. (5) with $\tau_{esc}= 0.5$ 
Gyr during the first 2 Gyr of galactic evolution and $\tau_{esc}=\infty$ afterwards 
(i.e no time variation of $\Lambda$). This means that the CR escape time-scale at 
$t=0$ was a factor $\sim 500$ higher than now.

\section{Results and conclusions}

Figures 1 and 2 show the calculated time-dependent GCR fluxes as well as the 
corresponding LiBeB evolution with [Fe/H]. The results are:
 
{\bf Case 1:} The flux is {\it higher in the past} and decreases slowly with time in the 
whole energy range. This higher flux implies a larger LiBeB production in the past than 
the SM model, although, this is still not enough to reproduce the observations. 

{\bf Case 2:} The flux in the past is {\it also higher and the spectrum is flatter at 
energies E$\geq 100$} MeV/n. In consequence the LiBeB production is even higher at early 
epochs than in {\it Case 1}. Note that the predicted Be evolution is nearly in 
agreement with the observations. 

\begin{figure}
\centerline
{\vbox{\psfig{figure=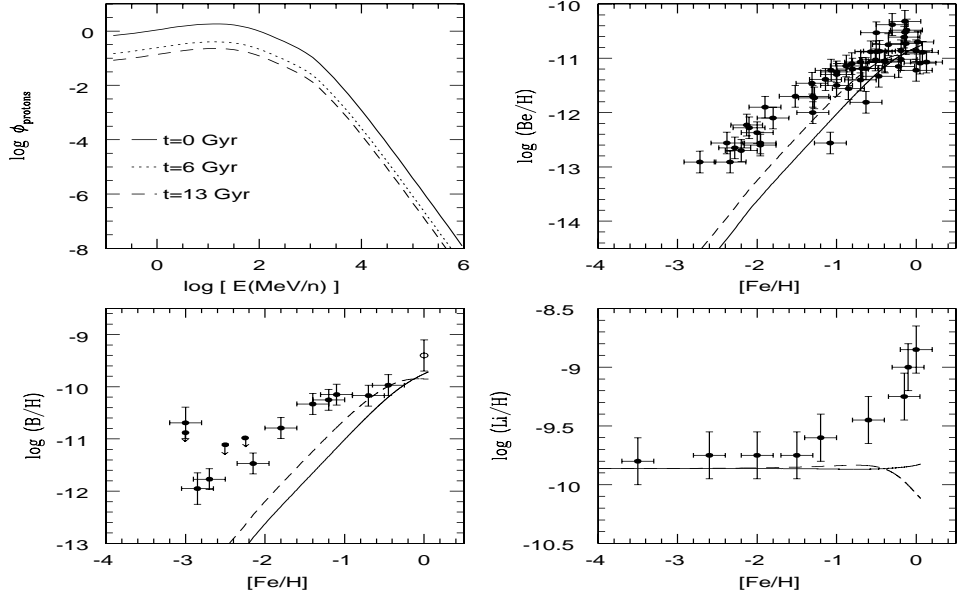,height=8.2 cm,width=\textwidth}}}
\caption[]{\small
Calculated GCR flux (upper left panel) and evolution of the LiBeB abundances in Case 1
(dashed line) compared with the predictions of the standard model (continuous line).
}
\end{figure}
\begin{figure}
\centerline
{\vbox{\psfig{figure=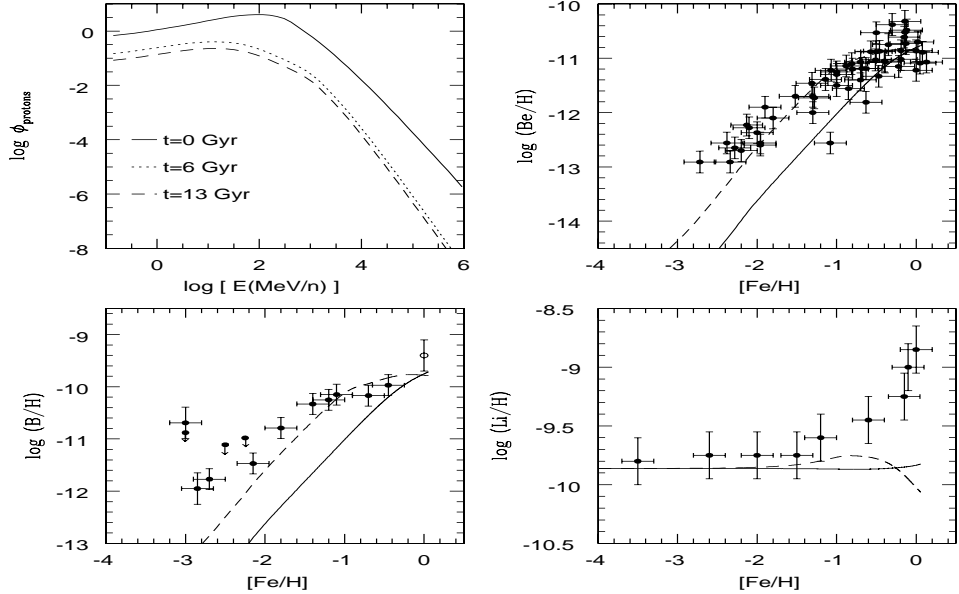,height=8.2 cm,width=\textwidth}}}
\caption[]{\small
Calculated GCR flux (upper left panel) and evolution of the LiBeB abundances in Case 2
(dashed line) compared with the predictions of the standard model (continuous line).
}
\end{figure}
On the other hand, the effect of considering the temporal evolution of the gas content
in the galaxy is not very important for the LiBeB production. In this case the flux
and spectrum are modified in a complex way depending on the value of $\tau_{gas}$ and the 
interplay between $\Lambda_{e}$ and $\Lambda_n$. Our results show that variations of 
the GCR flux in time have to be considered to study the early production of LiBeB by 
spallation in the ISM. A higher flux and a flatter CR spectrum in 
the past are able to improve the  agreement between the predictions of the (modified) 
SM with observations. This is evident for Be and B. For Li, however, another important 
source of production is still required. A higher flux in the past can be achieved by 
taking into account the temporal evolution of the GCR sources, which is naturally 
explained by the temporal variation of the SN rate, or by assuming that the escape 
time-scale $\Lambda_{esc}$ was higher in the past. The later assumption can improve 
the agreement between data and model considerably. However, it still requires a more 
solid theoretical justification (\cite{Pr}).  

\vfill
\end{document}